\documentclass[letterpaper, 10 pt, conference]{ieeeconf}  

\IEEEoverridecommandlockouts                              
\usepackage{amsmath}
\usepackage{amsfonts}
\usepackage{amssymb}
\usepackage{color}
\usepackage{soul}
\usepackage{cite}
\usepackage{graphicx}
\usepackage{algorithm}
\usepackage[noend]{algpseudocode}
\usepackage{algorithmicx}
\usepackage{mathtools}
\usepackage{bm}
\usepackage{booktabs}
\usepackage{pgf}
\usepackage[normalem]{ulem} 

\usepackage{caption}
\usepackage{subcaption}
\usepackage[colorlinks=false, allcolors=blue]{hyperref}
\usepackage{adjustbox}
\usepackage{makecell}

\DeclareMathSymbol{\shortminus}{\mathbin}{AMSa}{"39}

\usepackage{booktabs}

\algnewcommand\algorithmicswitch{\textbf{switch}}
\algnewcommand\algorithmiccase{\textbf{case}}
\algnewcommand\algorithmicassert{\texttt{assert}}
\algnewcommand\Assert[1]{\State \algorithmicassert(#1)}%

\algdef{SE}[SWITCH]{Switch}{EndSwitch}[1]{\algorithmicswitch\ #1\ \algorithmicdo}{\algorithmicend\ \algorithmicswitch}%
\algdef{SE}[CASE]{Case}{EndCase}[1]{\algorithmiccase\ #1}{\algorithmicend\ \algorithmiccase}%
\algtext*{EndSwitch}%
\algtext*{EndCase}%

\newcommand{\eye}{\mathbf{I}}
\newcommand{\zeros}{\mathbf{0}}

\newcommand{\norm}[1]{\left\lVert #1 \right\rVert}

\newcommand{\Reals}{\mathbb{R}}

\newcommand{\xx}{\mathbf{x}}
\newcommand{\uu}{\mathbf{u}}
\newcommand{\xxi}{\bm{\xi}}
\newcommand{\ff}{\mathbf{f}}
\newcommand{\ww}{\mathbf{w}}
\newcommand{\vv}{\mathbf{v}}
\newcommand{\yy}{\mathbf{y}}
\newcommand{\zz}{\mathbf{z}}
\newcommand{\ee}{\mathbf{e}}
\newcommand{\llambda}{\bm{\lambda}}
\newcommand{\cc}{\mathbf{c}}
\newcommand{\bb}{\mathbf{b}}
\newcommand{\qq}{\mathbf{q}}
\newcommand{\clint}[2]{\left[{#1},{#2}\right]}

\newcommand{\StateSet}{\mathcal{S}}

\title{\LARGE \bf
Moving Horizon Estimation of Hybrid Systems Using Hybrid Zonotopes and Mixed-Integer Quadratic Programs}

\author{
Jonah J. Glunt and Herschel C. Pangborn
\thanks{Authors are with the Department of Mechanical Engineering, The Pennsylvania State University, University Park, PA 16802 USA (e-mail: {\tt\small jglunt@psu.edu, hcpangborn@psu.edu}).}
\thanks{This work was supported by the Office of Naval Research under Award N000142512051. Any opinions, findings, and conclusions or recommendations expressed in this material are those of the authors and do not necessarily reflect the views of the Office of Naval Research.}
}
\begin{document}

\maketitle
\thispagestyle{empty}
\pagestyle{empty}

\begin{abstract}
Moving horizon estimation (MHE) is an estimation technique that formulates an optimization problem over a past horizon to find the states that require the minimum disturbances and noise to be feasible with the dynamics and measurements. For hybrid systems with both continuous and discrete states, this yields a mixed-integer program for the MHE, which can be difficult both to formulate and solve online under computational limitations. 
This paper shows how the hybrid zonotope representation for mixed-integer sets can be used to efficiently construct the feasible space for the program using reachability-based calculations for piecewise affine systems, without requiring conversion to a mixed logical dynamical system. 
A method for lifting the state space to explicitly estimate the dynamic mode is also presented. 
The proposed approach is evaluated in two numerical examples. 
Comparison against a standard mixed logical dynamical systems approach shows the proposed method can be solved orders of magnitude faster for one of the example problems. 
\end{abstract}

\section{Introduction}
\subsection{Motivation}
State estimation is a fundamental challenge in control of hybrid systems, where both continuous states and discrete modes must be determined from noisy measurements. 
Moving horizon estimation (MHE) is an optimization-based framework that yields a state estimate by minimizing the dynamic disturbances and measurement noises required for measurements to be feasible with the modeled dynamics and initial condition. 
Performing MHE on hybrid systems is computationally challenging, as it results in mixed-integer optimization programs that are NP-hard. 
Recent work~\cite{robbins2026hybrid} has presented efficient methods for reachability of PWA systems as well as a heuristic for mixed-integer quadratic programs over hybrid zonotopes, which this paper applies toward MHE. 

\subsection{Literature Review}
MHE for linear and nonlinear systems has been well studied in the literature. For an unconstrained linear system, a Kalman filter or other linear estimator can be used~\cite{muske1995nonlinear}, and in the presence of constraints MHE results in a convex program~\cite{rao2001constrained} or alternatively a constrained Kalman-type approach~\cite{Simon2010_constrained_KF}. MHE for smooth nonlinear systems has also been examined and guarantees for convergence and stability have been derived~\cite{Michalska_nonlinMHE, rao2003constrained, Muller2017_nonlinMHE, Knuffer2023_nonlinMHE}.

As hybrid systems involve discrete modes or decision variables, MHE formulations for hybrid systems result in mixed-integer programs. Foundational papers on MHE of hybrid systems are~\cite{Bemporad1999_hybridMHE}, which formulates a mixed-integer quadratic program (MIQP), and~\cite{FerrariTrecate2002_hybridMHE}, which provides conditions for asymptotic convergence of the MIQPs in a moving horizon. This general formulation has been extended to a min-max setting for robustness to uncertainty~\cite{Rowe2003_hybridMHE} and to estimating unknown inputs or disturbances~\cite{Pina2006_hybridMHE}. These papers all use the mixed logical dynamical (MLD) description of hybrid systems, which is equivalent to piecewise affine (PWA) systems under well-posed conditions. 

An important subclass of PWA and MLD systems are switched systems, which have also seen explicit attention in the MHE literature. 
In~\cite{Alessandri2005_switchMHE}, systems with unrestricted and unknown switching between linear modes are considered, and observability and convergence results are derived. 
These have been extended to nonlinear switched systems~\cite{Guo2013_switchMHE}, with the mode treated as an unknown discrete state. 

Outside of MHE, more generic set-valued state estimation approaches have been developed for various set representations, such as ellipsoids~\cite{Chernousko2005_ellipsoidal_estimation, Chabane2014_ellipsoid_estimation}, polytopes~\cite{Wan2018_nonlin_est_polytopes}, constrained convex generators~\cite{Silvestre2023_SVSE_CCG}, and zonotopic representations (zonotopes, constrained zonotopes, and hybrid zonotopes). This paper focuses on zonotopic set representations because they are closed under the set operations needed for reachability analysis and estimation~\cite{althoff2021set} and to utilize efficient solvers written to exploit their structure~\cite{robbins2026hybrid}. Zonotopes and constrained zonotopes have been applied to perform set-valued estimation of linear~\cite{scott2016constrained,pan2022moving,Wagner2026}, nonlinear~\cite{Alamo2005, Combastel2005AZonotopes ,rego2020guaranteed}, and hybrid systems~\cite{Tabatabaeipour2013_zono_PWA, Zhang2020_zono_switched_est, Fei2022_zono_switch_est}. Hybrid zonotopes specifically have been used for estimation of nonlinear systems using approximation techniques~\cite{siefert2023set,thompson2025mixed}, but not for hybrid systems by exactly representing their dynamics. 

\subsection{Contributions}
This paper considers the problem of moving horizon estimation for hybrid systems. The formulation of the MHE problem for PWA systems as an MIQP over a hybrid zonotope feasible set is detailed, without requiring conversion to an MLD system description. Methods are provided for the construction of the hybrid zonotope feasible space that encodes all consistent state trajectories and mode sequences over the horizon. Further, mode sequence recovery is covered as an explicit part of the estimate via state-space lifting with mode indicator variables. This paper includes some discussion of practical considerations of the moving horizon implementation and periodic complexity reduction of the estimated state sets via over-approximation. The proposed method is shown to be solved orders of magnitude faster on an example problem than the mixed logical dynamical approach standard in the literature. 

\subsection{Outline}
The remainder of this paper is organized as follows. Section~\ref{sec:prelims} describes the notation and preliminary material, including hybrid zonotopes and the class of hybrid systems in consideration. Section~\ref{sec:methods} presents the proposed method for formulating the MHE problem as a MIQP over a hybrid zonotope. Two numerical examples are given in Section~\ref{sec:examples}, and Section~\ref{sec:conclusion} concludes the paper. 
\section{Preliminaries}
\label{sec:prelims}

\subsection{Notation and General Definitions}
Unless otherwise stated, scalars are denoted by lowercase letters, vectors by boldface lowercase letters, matrices by uppercase letters, and sets by calligraphic letters. 
The symbol $\zeros$ denotes a vector or matrix consisting entirely of zeros, with dimensions provided in subscript when needed.  
The identity matrix is denoted $\eye$, again with dimension in subscript when not clear from context.
The closed interval $\clint{\mathbf{a}}{\mathbf{b}} = \{\mathbf{x}\mid \mathbf{a}\leq \mathbf{x} \leq \mathbf{b}\}\subset\mathbb{R}^n$, where $\mathbf{a},\mathbf{b}\in\mathbb{R}^n$. 
The concatenation of two vectors is notated by $(\xx,\yy) = \begin{bmatrix} \xx^T & \yy^T \end{bmatrix}^T$. 
The weighted vector norm $\norm{\xx}_Q^2=\xx^T Q \xx$. 

Given a function $\mathbf{f}: \mathcal{X}\rightarrow\mathcal{Y}$, the \emph{graph of the function} is the set
\begin{equation}
    \label{eqn:GoF_defn}
    \left\{ (\xx,\yy) \mid \xx\in\mathcal{X}\;,\ \yy = \mathbf{f}(\xx) \right\}\;.
\end{equation}

\subsection{Piecewise Affine Hybrid System Model}

This work considers \emph{piecewise affine} (PWA) hybrid systems of the form
\begin{subequations}
\label{eqn:PWA_system_defn}
\begin{align}
    \xx_{k+1} &= A_k^i \xx_k + B_k^i \uu_k + \ff_k^i+\ww_k\;,\\
    (\xx_k, \uu_k) &\in \mathcal{D}_k^i\;,\\
    \yy_k &= C_k \xx_k + \vv_k\;,\\
    \ww_k &\in \mathcal{W}\;,\; \vv_k \in \mathcal{V}
\end{align}
\end{subequations}
where the dynamic mode $i\in\{1,\dots,n_m\}$ depends on constraints coupled between the states $\xx_k\in\Reals^{n_x}$ and input $\uu_k\in\Reals^{n_u}$, written as the set $\mathcal{D}_k^i\subseteq\Reals^{n_x+n_u}$, but the model for outputs $\yy_k\in\Reals^{n_y}$ does not vary by mode. The states are affected by unknown but bounded additive process noise $\ww_k$, and similarly the outputs by measurement noise $\vv_k$, neither of which depend on the dynamic mode. This PWA form can represent switched systems by lifting the state space to include mode indicator dimensions, either as states or inputs. The methods presented in this paper can be extended to handle additional modal couplings with the noises and outputs.

We further consider the state space to be bounded in some set $\mathcal{S}$ and the inputs to come from the set $\mathcal{U}$. We assume that $\cup_i^{n_m} \mathcal{D}_k^i = \mathcal{S} \times \mathcal{U}$.

\subsection{Moving Horizon Estimation}
Given a dynamic model for the evolution of states and outputs of a system, \emph{moving horizon estimation} (MHE) 
seeks to find an optimal state trajectory that minimizes state disturbances and measurement noises consistent with the modeled system dynamics and observed (noisy) measurements.
An optimization program is solved for each call of the estimator in a moving horizon fashion, meaning it operates over a sliding window of recent measurements. This is the dual problem of the optimal control framework \emph{model predictive control}. 

Mathematically, given the measurements $\{\yy_k\}_{k=1}^{N}$ and inputs $\{\uu_k\}_{k=0}^{N-1}$, the general MHE problem is defined as follows:
\begin{subequations} \label{eqn:MHE_general}
\begin{align}
    \min & \norm{\xx_0-\xx^r_0}^2_P + \sum_{k = 0}^{N-1} \norm{\ww_k}^2_Q + \norm{\vv_{k+1}}^2_R \\
    \text{s.t. } & \forall k \in \{0,\dots,N-1\}:\\
    &\quad \xx_{k+1} = \mathbf{f}(\xx_k, \uu_k, \ww_k) \;,\\
    &\quad \yy_{k+1} = \mathbf{g}(\xx_{k+1}, \vv_{k+1})\;,\\
    &\quad \xx_{k+1} \in \StateSet\;,\ \ww_k \in \mathcal{W}\;,\ \vv_{k+1} \in \mathcal{V}\;,\\
    &\xx_0 \in \mathcal{X}_0\;.
\end{align}
\end{subequations}
The \emph{horizon} is $N$ steps, with the nomenclature that at each call of the MHE, $\xx_0$ is the state furthest in the past and $\xx_N$ is the current state. The set $\mathcal{X}_0$ is a hard constraint on the initial set, taken as the estimated set at that time from the previous MHE iteration, or a known bounding set around the initial condition at startup. The term $\norm{\xx_0-\xx_0^r}_P^2$ in the cost function is the \emph{arrival cost} which penalizes deviation of the oldest state in the horizon from a prior reference $\xx_0^r$, typically taken from the optimal estimate of the previous iteration of the MHE. The weighting matrices $P$, $Q$, and $R$ are design parameters to balance uncertainty in the prior estimate with minimizing disturbances and noises over the current horizon. The matrix $P$ can also be selected for guaranteed convergence or stability~\cite{FerrariTrecate2002_hybridMHE}.

Using the PWA dynamics in~\eqref{eqn:PWA_system_defn} and substituting in $\yy_{k+1} = C_{k+1} \xx_{k+1} + \vv_{k+1}$, the MHE optimization problem can be rewritten as
\begin{subequations} \label{eqn:MHE_PWA}
\begin{align}
    \min & \norm{\xx_0-\xx^r_0}^2_P + \sum_{k = 0}^{N-1} \norm{\ww_k}^2_Q + \norm{\xx_{k+1}}^2_{C^T R C} \nonumber\\
    &\qquad\qquad\qquad\qquad- 2\yy_{k+1}^T R C \xx_{k+1} \\
    \text{s.t. } & \forall k \in \{0,\dots,N-1\}: \nonumber\\
    &\quad \xx_{k+1} = A_k^i \xx_k + B_k^i \uu_k + \ff_k^i + \ww_k \nonumber\\
    &\qquad\qquad\text{ if } (\xx_k,\uu_k)\in\mathcal{D}_k^i \;,\\
    &\quad \xx_{k+1} \in \StateSet\:,\ww_k \in \mathcal{W}\:,\: \yy_{k+1}-C_{k+1}\xx_{k+1} \in \mathcal{V}\:,\\
    &\xx_0 \in \mathcal{X}_0\;.
\end{align}
\end{subequations}
This allows $\yy$ and $\vv$ to be eliminated as decision variables from the optimizer and the problem can be written only in terms of $\xx$ and $\ww$, while still using the cost function designed for the noise. 

\subsection{Zonotopic Set Representations}
A set $\mathcal{Z}$ is a zonotope if $\exists G \in \Reals^{n \times n_G}$ and $\cc \in \Reals^n$ such that~\cite{girard2005reachability}
\begin{equation}
    \label{eqn:zono_defn}
    \mathcal{Z} = \left\{ G \xxi + \cc ~\middle|~ \xxi \in [-1,1]^{n_G} \right\} \;.
\end{equation}
Zonotopes can represent any centrally symmetric convex polytopes.

Constrained zonotopes extend the definition of zonotopes to include equality constraints. A set $\mathcal{Z}$ is a constrained zonotope if there exists a generator matrix $G \in \Reals^{n \times n_G}$, center $\cc \in \Reals^n$, constraint matrix $A \in \Reals^{n_C \times n_G}$, and constraint vector $\bb \in \Reals^{n_C}$ such that
\begin{equation}\label{eqn:conzono_defn}
    \mathcal{Z} = \left\{ G \xxi + \cc ~\middle|~ A \xxi = \bb,\; \xxi \in [-1,1]^{n_G} \right\} \;.
\end{equation}
Constrained zonotopes can represent any convex polytope \cite{scott2016constrained}.

Hybrid zonotopes extend constrained zonotopes by allowing for both continuous and binary factors, $\xxi_c$ and $\xxi_b$. A set $\mathcal{Z}$ is a hybrid zonotope if there exist continuous and binary generator matrices $G_c \in \Reals^{n \times n_{Gc}}$ and $G_b \in \Reals^{n \times n_{Gb}}$, center $\cc \in \Reals^n$, continuous and binary constraint matrices $A_c \in \Reals^{n_C \times n_{Gc}}$ and $A_b \in \Reals^{n_C \times n_{Gb}}$, and constraint vector $\bb \in \Reals^{n_C}$ such that
\begin{equation}\label{eqn:hybzono_defn}
    \mathcal{Z} = \left\{ G_c \xxi_c + G_b \xxi_b + \cc ~\middle|~\begin{matrix} 
        A_c \xxi_c + A_b \xxi_b = \bb\;, \\ 
        \xxi_c \in [-1,1]^{n_{Gc}}\;,\\ 
        \xxi_b \in \{-1, 1\}^{n_{Gb}} 
    \end{matrix} \right\} \;.
\end{equation}
Hybrid zonotopes can represent any finite union of bounded convex polytopes~\cite{bird2023hybrid}.

In the above definition, $n$ is the dimension of the hybrid zonotope, $n_{Gc}$ is the number of continuous generators, $n_{Gb}$ is the number of binary generators, $n_G = n_{Gc} + n_{Gb}$ is the total number of generators, and $n_C$ is the number of equality constraints. The complexity of a hybrid zonotope refers to the number of continuous generators, binary generators, and constraints it has, i.e., $(n_{Gc}, n_{Gb}, n_C)$.

One major advantage of zonotopic sets when compared to other set representations is that they have closed form expressions for key set operations. 
In particular, hybrid zonotopes are closed under affine maps, Cartesian products, Minkowski sums, and generalized intersections. Identities for these operations on hybrid zonotopes can be found in~\cite{bird2023hybrid}.
\section{Hybrid Zonotope Formulation of MHE for PWA Systems}
\label{sec:methods}
This section details how to formulate the MHE problem for PWA systems using hybrid zonotopes. First, Section~\ref{subsec:GoF} reviews how a hybrid zonotope can represent the graph of a PWA function for the dynamics. Then, Section~\ref{subsec:feasbile_space} describes how to construct the feasible space given the graphs of the time-varying dynamics over the horizon, the received measurements, and the applied inputs. Section~\ref{subsec:mode_lifting} shows how the state space can be lifted to explicitly estimate the discrete dynamic mode at each time step. 
Lastly, Section~\ref{subsec:complexity_reduction} explains how the feasible sets are propagated from one MHE call to another and methods for performing periodic complexity reduction.

\subsection{Graph of Dynamics Function}
\label{subsec:GoF}
Combining the definitions for the graph of a function~\eqref{eqn:GoF_defn} and a PWA system~\eqref{eqn:PWA_system_defn}, we obtain the graph of PWA system dynamics as
\begin{align}
    \Psi_k& = \big\{ (\xx_k, \uu_k, \xx_{k+1}) \in \StateSet \times \mathcal{U} \times \Reals^{n_x} \;| \nonumber\\
    &\xx_{k+1} = A_k^i \xx_k + B_k^i \uu_k + \ff_k^i \text{ if } (\xx_k, \uu_k) \in \mathcal{D}_k^i \big\}\;.
\end{align}
The state disturbances $\ww_k \in \mathcal{W}$ are not considered in $\Psi_k$ because they are assumed to be mode-independent and will be handled in a different manner. 

Hybrid zonotopes can exactly represent the graph of PWA system dynamics~\cite{robbins2026hybrid}. This can be done by representing the graph of each affine piece and then taking the union over all pieces. Specifically, the graph of the affine dynamics over the domain piece $\mathcal{D}_k^i$ can be constructed as
\begin{align}
    \Psi_k^i = \begin{bmatrix}
        \eye_{n_x+n_u} \\
        \begin{matrix}
            A_k^i & B_k^i
        \end{matrix}
    \end{bmatrix} \mathcal{D}_k^i + \begin{bmatrix} \zeros \\ \ff_k^i \end{bmatrix}\;,
\end{align}
and then the overall graph is constructed by taking the union over each piece, i.e., 
\begin{align}
    \Psi_k = \bigcup_{i=1}^{n_m} \Psi_k^i\;,
\end{align}
where the union is performed either using the identity in~\cite{glunt2025sharp} to preserve sharpness of the convex relaxation, or using one of the identities from~\cite{robbins2026hybrid} for reduced complexity.

\subsection{Constructing Hybrid Zonotope Feasible Space}
\label{subsec:feasbile_space}
To build a hybrid zonotope representation of the feasible space for MHE written in the form~\eqref{eqn:MHE_PWA}, a hybrid zonotope $\mathcal{Z}_N$ is constructed such that any point $\zz\in\mathcal{Z}_N$ satisfies:
\begin{subequations}
\begin{align}
    \zz &= (\xx_0, \ww_0, \dots, \xx_{N-1}, \ww_{N-1}, \xx_N)\;,\\
    \xx_0 &\in \mathcal{X}_0\;,\\
    \forall k&\in\{0, \dots, N-1\}:\nonumber\\
    &\quad \ww_k \in \mathcal{W} \;,\\
    &\quad \xx_{k+1} \in \StateSet \;,\\
    &\quad \exists \vv_{k+1}\in\mathcal{V} \;,\; \yy_{k+1} = C_{k+1} \xx_{k+1} + \vv_{k+1} \;,\\
    &\quad (\xx_k, \uu_k, \xx_{k+1}-\ww_k) \in \Psi_k \;,
\end{align}
\end{subequations}
where $\mathcal{X}_0$ is the set of states for the earliest step in the horizon (from the previous iteration of MHE), and $\Psi_k$ are the graphs of the dynamics function at each step in the horizon. If the dynamics are not time-varying, then $\Psi_k$ can be constant across time steps. 

The set $\mathcal{Z}_N$ is constructed by first setting $\mathcal{Z}_0=\mathcal{X}_0$ and then recurring
\begin{subequations}
\begin{align}
    \bar{\mathcal{X}}_{k+1} &= \mathcal{S} \cap_{C_{k+1}} \left( \yy_{k+1} + (-\mathcal{V}) \right)\;,\\
    \Psi_k' &= M \left( \Psi_k \cap_{R_1} \{\uu_k\}\right)\;,\\
    \mathcal{Z}_{k+1} &= \left( \mathcal{Z}_k \times \mathcal{W} \times \bar{\mathcal{X}}_{k+1} \right) \cap_{R_{2,k}} \Psi_k'\;,
\end{align}
\end{subequations}
for $k=0,\dots,N-1$, where
\begin{subequations}
\begin{align}
    R_{1} &= \begin{bmatrix}
        \zeros_{n_u \times n_x} & \eye_{n_u} & \zeros_{n_u \times n_x} 
        \end{bmatrix} \;,\\
    R_{2,k} &= \begin{bmatrix}
        \zeros_{2n_x \times 2 k n_x} & \begin{matrix}
            \eye_{n_x} & \zeros & \zeros\\
            \zeros & -\eye_{n_x} & \eye_{n_x}
        \end{matrix}
    \end{bmatrix} \;,\\
    M &= \begin{bmatrix}
         \eye_{n_x} & \zeros_{n_x\times n_u} & \zeros \\
         \zeros & \zeros & \eye_{n_x}
    \end{bmatrix} \;.
\end{align}
\end{subequations}
The set $\bar{\mathcal{X}}_{k+1}$ represents the set of states at time $k+1$ that are consistent with the measurement received, and $\Psi_k'$ is the graph of the dynamics function with the known applied input enforced and the input dimension projected away. The new iterate $\mathcal{Z}_{k+1}$ is constructed by first performing a Cartesian product of the previous iterate with the disturbance set and the set of measurement consistent states, then intersecting with the graph of the dynamics (with the applied input enforced).

\subsection{Lifting State Dimension for Mode Tracking}
\label{subsec:mode_lifting}

To examine the mode trajectory associated with the optimal estimate (or feasible modes associated with feasible state sets), then the state dimension can be lifted to include an indicator variable for each mode. The lifted state is written as $\tilde{\xx} = (\xx, \llambda)\in\StateSet\times\{0,1\}^{n_m}$, where $\sum\lambda_i = 1$ and $\lambda_i=1$ when the system is in mode $i$. The rest of the constructions are modified as follows. 

The lifted graph of mode $i$'s dynamics at time step $k$ is defined as
\begin{align}
    \tilde{\Psi}_k^i = \big\{(\xx_k,\uu_k,&\llambda_k,\xx_{k+1}) \in \tilde{\mathcal{D}}_k^i \times \Reals^{n_x} \;| \nonumber\\
    &\xx_{k+1} = A_k^i \xx_k + B_k^i \uu_k + \ff_k^i \big\}\;,
\end{align}
where $\tilde{\mathcal{D}}_k^i = \mathcal{D}_k^i\times\{\ee_i\}$ and $\ee_i$ is the $i$-th column of the identity matrix $\eye_{n_m}$.
Each $\tilde{\Psi}_k^i$ can be constructed via
\begin{equation}
    \tilde{\Psi}_k^i = \begin{bmatrix}
        {} & \eye_{n_x+n_u+n_m} & {} \\ 
        A_k^i & B_k^i & \zeros 
    \end{bmatrix} \tilde{\mathcal{D}}_k^i + \begin{bmatrix}
        \zeros \\ \ff_k^i
    \end{bmatrix}\;.
\end{equation}

Then, as before, the overall graph of the dynamics is constructed by taking the union over each piece, i.e., $\tilde{\Psi}_k = \cup_i \tilde{\Psi}_k^i$. 

Accordingly, construction of the MHE feasible set is also modified for the lifted dimensions. The feasible space construction is initiated with $\tilde{\mathcal{Z}}_0=\tilde{\mathcal{X}}_0$, with the lifted initial condition $(\xx_0, \llambda_0) \in \tilde{\mathcal{X}}_0$, where $\tilde{\mathcal{X}}_0 \subseteq \mathcal{S}\times\mathcal{I}$, where $\mathcal{I}=\{0,1\}^{n_m}$ represents the $n_m$ lifted states for the mode selection. The lifted form recursion can then be written
\begin{subequations}
\begin{align}
    \bar{{\mathcal{X}}}_{k+1} &= \mathcal{S} \cap_{C_{k+1}} (\yy_{k+1} + (-\mathcal{V}))\;,\\
    \tilde{\Psi}_k' &= \tilde{M} \left(\tilde{\Psi}_k \cap_{\tilde{R}_1} \{\uu_k\}\right)\;,\\
    \tilde{\mathcal{Z}}_{k+1} &= \left( \tilde{\mathcal{Z}}_k \times \mathcal{I} \times \mathcal{W} \times \bar{{\mathcal{X}}}_{k+1} \right) \cap_{\tilde{R}_{2,k}} \tilde{\Psi}_k'\;,
\end{align}
\end{subequations}
for $k=0,\dots,N-1$, where
\begin{subequations}
\begin{align}
    \tilde{R}_1 &= \begin{bmatrix}
        \zeros_{n_u \times n_x} & \eye_{n_u} & \zeros_{n_u \times n_m} & \zeros_{n_u \times n_x}
    \end{bmatrix}\;,\\
    \tilde{R}_{2,k} &= \begin{bmatrix}
        \zeros_{(2n_x +n_m) \times k(2n_x + n_m)} & \begin{matrix}
            \eye_{n_x} & \zeros & \zeros & \zeros \\
            \zeros & \eye_{n_m} & \zeros & \zeros \\
            \zeros & \zeros & -\eye_{n_x} & \eye_{n_x}
        \end{matrix}
    \end{bmatrix}\;,\\
    \tilde{M} &= \begin{bmatrix}
        \eye_{n_x} & \zeros_{n_x\times n_u} & \zeros & \zeros \\
        \zeros & \zeros & \eye_{n_m} & \zeros \\
        \zeros & \zeros & \zeros & \eye_{n_x}
    \end{bmatrix}\;.
\end{align}
\end{subequations}
In practice, to eliminate unnecessary binary variable complexity, if the mode is uncertain when defining the initial condition $\tilde{\mathcal{X}}_0$, then continuous intervals $\clint{0}{1}$ can be used instead of the discrete $\{0,1\}$ in $\mathcal{I}$, since the discrete switching logic is exactly represented in $\tilde{\Psi}_k$. 

Assuming that $\mathcal{S}$, $\mathcal{W}$, $\mathcal{V}$ are all zonotopes with $n_G^\mathcal{S}$, $n_G^\mathcal{W}$, $n_G^\mathcal{V}$ generators each, and assuming $\Psi$ is time-independent and has hybrid zonotope complexity $(n_{Gc}^\Psi, n_{Gb}^\Psi, n_C^\Psi)$, using continuous variables for $\mathcal{I}$, and given that the set $\mathcal{X}_0$ has hybrid zonotope complexity $(n_{Gc}^{\mathcal{X}_0}, n_{Gb}^{\mathcal{X}_0}, n_C^{\mathcal{X}_0})$, then the complexity of $\tilde{\mathcal{Z}}^N$ is
\begin{subequations}
\begin{align}
    n_{Gc} &= n_{Gc}^{\mathcal{X}_0} + N \left( n_G^\mathcal{S} + n_G^\mathcal{W} + n_G^\mathcal{V} + n_{Gc}^\Psi +n_m\right) \;,\\
    n_{Gb} &= n_{Gb}^{\mathcal{X}_0} + N n_{Gb}^\Psi\;,\\
    n_C &= n_C^{\mathcal{X}_0} + N \left( 2n_x + n_y + n_u + n_C^\Psi +n_m\right)\;.
\end{align}
\end{subequations}
If not using the lifted form, then the $n_m$ terms are simply dropped. 

\subsection{Moving Horizon Propagation and Periodic Complexity Reduction}
\label{subsec:complexity_reduction}
Historical data from before the horizon is considered in the MHE by propagating the initial condition sets. After solving MHE at one time step, the initial condition set $\mathcal{X}_0$ or $\tilde{\mathcal{X}}_0$ for the next time step can be projected out of the previous feasible set $\mathcal{Z}_N$ or $\tilde{\mathcal{Z}}_N$, and then used to seed the feasible set construction. However, projecting the new $\mathcal{X}_0$ out of the old $\mathcal{Z}_N$ retains the full complexity of the feasible space in the initial condition set, and therefore the representation complexity for feasible space $\mathcal{Z}_N$ grows with each call of MHE. Therefore, some periodic complexity reduction is required to keep the complexities tolerable for solving online. 

Typically, the periodic complexity reduction is performed via over-approximation with a set of simpler complexity. The method that yields the lowest possible complexity would be to take the interval hull, also called bounding box, of the initial state set (and modes, if in the lifted form), which can be accomplished via $2 n_x$ (or $2(n_x+n_m)$) mixed-integer linear program support functions. However, in the lifted case, this eliminates all coupling between the state estimates and mode estimates. Another method is to take the convex relaxation of the hybrid zonotope representing $\mathcal{X}_0$, which yields a constrained zonotope over-approximation, and then perform constrained zonotope simplification techniques~\cite{scott2016constrained,Dulal2026_CZ_reduction}. A third option is to take the union of interval hulls corresponding to each dynamic mode, rather than an interval hull over all states and modes. This preserves some coupling between states and modes, at the expense of increased complexity over the bounding box. 
\section{Numerical Examples}
The following examples were performed using the proposed methods from the previous section. The MHE problems were implemented in Python and the resulting MIQPs were solved using the ADMM-based heuristic presented in~\cite{robbins2026hybrid}. Timing data was obtained on an Ubuntu 24.04 desktop computer with an Intel\textregistered~Core\textsuperscript{\tiny TM} Ultra 7 265 × 20 processor and 32~GB of RAM.

\label{sec:examples}
\subsection{Lane-changing Vehicle}
This example demonstrates the proposed approach on the lane-changing vehicle control example presented in~\cite{robbins2026hybrid}. Consider double integrator vehicle dynamics in road-aligned Frenet coordinates $s$ (distance along the road center line) and $d$ (perpendicular to road center line, positive in right lane) given as $\xx_{k+1} = A\xx_k + B\uu_k + \ww_k$  where
\begin{subequations}\label{eqn:double_integrator}
\begin{align}
    \xx_k &= (s_k, d_k, \dot{s}_k, \dot{d}_k)\;,\\
    \uu_k &= (\ddot{s}_k, \ddot{d}_k)\;,\\
    A &= \begin{bmatrix}
        1 & 0 & \Delta t & 0 \\
        0 & 1 & 0 & \Delta t \\
        0 & 0 & 1 & 0 \\
        0 & 0 & 0 & 1
    \end{bmatrix}\;,\  B = \begin{bmatrix}
        \frac{\Delta t^2}{2} & 0 \\
        0 & \frac{\Delta t^2}{2} \\
        \Delta t & 0 \\
        0 & \Delta t
    \end{bmatrix}\;,
\end{align}
\end{subequations}
and $\Delta t = 1$s. A lane-tracking controller of the form
\newcommand{\kff}{k_\mathit{ff}}
\begin{align}
\label{eq:lanechange_feedback}
    \ddot{d}_k = -k_d d_k -k_{\dot{d}}\dot{d}_k+\kff+\ddot{d}'
\end{align}
is assumed for lateral vehicle control, where $k_d=0.213$ 1/s$^2$ and $k_{\dot{d}}=0.652$ 1/s are feedback terms, $k_\textit{ff}$ is a feedforward term, and $\ddot{d}'$ is a small perturbation. Defining 
\begin{equation}
K = \begin{bmatrix} 0 & 0 & 0 & 0 \\
                    0 & k_d & 0 & k_{\dot{d}}
\end{bmatrix} \;,
\end{equation}
the affine dynamics for right and left lane tracking are
\begin{equation} \label{eq:closed-loop-lane-tracking-modes}
    \mathbf{x}_{k+1} = \begin{cases} A^{\mathit{rl}} \mathbf{x}_k + B^{\mathit{rl}} \mathbf{u}'_k + \mathbf{f}^{\mathit{rl}},& \text{right lane tracking,} \\
    A^{\mathit{ll}} \mathbf{x}_k + B^{\mathit{ll}} \mathbf{u}'_k + \mathbf{f}^{\mathit{ll}},& \text{left lane tracking,}
 \end{cases} 
\end{equation}    
where 
\begin{subequations}
\begin{align}
    &\mathbf{u}'_k = (\ddot{s}, \ddot{d}')\;, \\
    &A^{\mathit{rl}} = A^{\mathit{ll}} = A-BK \;, \\
    &B^{\mathit{rl}} = B^{\mathit{ll}} = B \;, \\
    &\mathbf{f}^{\mathit{rl}} = B K \begin{bmatrix} 0 & d^{\mathit{rl}} & 0 & 0 \end{bmatrix}^T \;, \\
    &\mathbf{f}^{\mathit{ll}} = B K \begin{bmatrix} 0 & d^{\mathit{ll}} & 0 & 0 \end{bmatrix}^T \;,
\end{align}
\end{subequations}
and $d^{\mathit{rl}}$ and $d^{\mathit{ll}}$ are the lateral positions of the center lines of the right and left lanes, respectively. Note that $\mathbf{f}_{\mathit{rl}}$ and $\mathbf{f}_{\mathit{ll}}$ correspond to the $k_{\mathit{ff}}$ term in~\eqref{eq:lanechange_feedback}. Additional mode-dependent constraints and interpretation of this system model can be found in~\cite{robbins2026hybrid}. 

\begin{figure}
    \centering
    \input{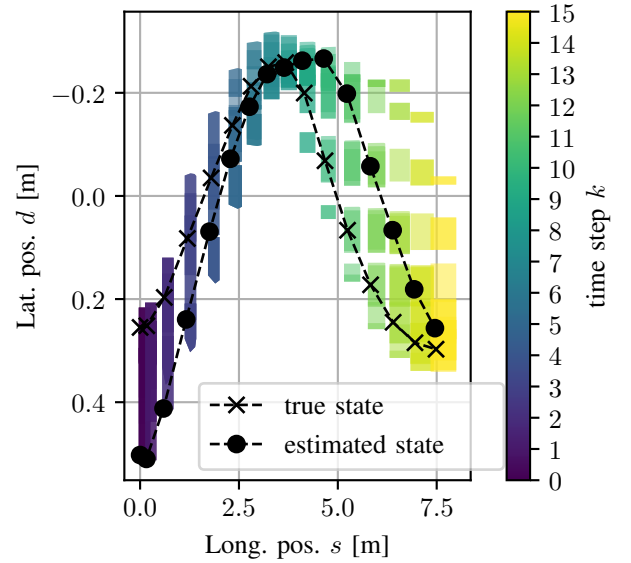}
    \caption{True states, estimated states, and feasible sets for the lane-changing vehicle MHE problem with horizon length $N=15$.}
    \label{fig:lanechange}
\end{figure}

For estimating this system, we consider a horizon of $N=15$ steps, with the measurement model $\yy_k = C \xx_k$, where $C = \begin{bmatrix} \eye_2 & \zeros \end{bmatrix}$, meaning the first two states (the position) are measured. The state disturbances $\ww_k$ are taken from $\mathcal{W} = \clint{-0.01}{0.01}^2 \times \clint{-0.001}{0.001}^2$, and the measurement noises $\vv_k$ from $\mathcal{V} = \clint{-0.75}{0.75} \times \clint{-0.5}{0.5}$. 

Figure~\ref{fig:lanechange} plots the true states, estimated states, and feasible sets for a single call of the estimator. At time steps toward the end of the horizon, uncertainty in the dynamic mode at earlier steps causes the feasible set to split into multiple disjoint polytopes, all represented by one hybrid zonotope. To evaluate the run-time performance of solving this mixed-integer MHE problem, 100 trials were timed. The median solution time was 26.2ms, with a minimum of 10.3ms and a maximum of 29.1ms.

\subsection{Two-room Thermal System}
\label{sec:thermal_example}
Consider a two-room thermal system with active heating, where the hot air supply to each room can be controlled by a physical vent, modeled as a discrete binary ``on/off". The estimator is unaware of the state of these vents, and therefore the system can operate under any one of four modes at each time step. Similar problems have been considered as a benchmark for hybrid systems verification~\cite{Fehnker2004_hybrid_benchmarks, Althoff2010_zonos_polytopes}.

The heater provides hot air with supply temperature $T_s$, specific heat $c_p$, and flow rate $\dot{m}$, which are assumed to be constant. The state of the heating vents leading to rooms one and two are described by $(\delta_1,\delta_2)\in\{0,1\}^2$, where $0$ indicates closed and $1$ indicates open. Room one has thermal capacitance $c_1$ and conductance $k_1$ with the outdoor temperature $T_\infty$; similarly $c_2$ and $k_2$ with room two. The conductance between the rooms is $k_{12}$. Table~\ref{tab:thermal_params} provides the parameter values used. 

\begin{table}
\centering
\caption{Two-room thermal system parameter values.}
\label{tab:thermal_params}
\begin{tabular}{lll}
\hline
Parameter & Value & Units \\
\hline
$c_p$      & 1005          & J/(kg$\cdot$K) \\
$c_1$      & $3\times10^6$ & J/K \\
$c_2$      & $3\times10^6$ & J/K \\
$k_1$      & 120           & W/K \\
$k_2$      & 120           & W/K \\
$k_{12}$   & 40            & W/K \\
$\dot{m}$  & 0.12          & kg/s \\
$T_s$      & 40            & $^\circ$C \\
$T_\infty$ & 0             & $^\circ$C \\
\hline
\end{tabular}
\end{table}

The dynamics are modeled by
\begin{subequations}
\label{eqn:2room_cont_dynamics}
\begin{align}
    c_1 \dot{T}_1 &= -k_1 (T_1-T_\infty) - k_{12} (T_1-T_2) +\delta_1 \dot{m} c_p (T_s-T_1)\;,\\
    c_2 \dot{T}_2 &= -k_2 (T_2-T_\infty) - k_{12} (T_2-T_1) + \delta_2 \dot{m} c_p (T_s-T_2)\;,
\end{align}
\end{subequations}
equivalently written $\dot{\xx} = A_c \xx + \ff_c$, with states $\xx=(T_1,T_2)$, where
\begin{subequations}
\begin{align}
    A_c &= \begin{bmatrix}
        -\frac{k_1+k_{12}+\delta_1\dot{m}c_p}{c_1} & \frac{k_{12}}{c_1} \\
        \frac{k_{12}}{c_2} & -\frac{k_2+k_{12}+\delta_2 \dot{m} c_p}{c_2}
    \end{bmatrix} \;,\\
    \ff_c &= \begin{bmatrix} \frac{k_1}{c_1} \\ \frac{k_2}{c_2} \end{bmatrix} T_\infty + \begin{bmatrix} \frac{\delta_1 \dot{m} c_p}{c_1} \\ \frac{\delta_2 \dot{m} c_p}{c_2} \end{bmatrix} T_s
\end{align}
\end{subequations}
The variables $T_s$, $\dot{m}$, $T_\infty$ could be treated as inputs or disturbances, but for this example, they are assumed constant. We consider the four combinations of $\bm{\delta}$ to obtain a switched linear system with four modes. This state space model is discretized exactly using a zero-order hold for the mode selection and a time step of 5 minutes. We will further assume that the discrete states experience additive disturbance $\ww$. We assume a mode-independent measurement matrix $C$ and additive measurement noise $\vv$, to obtain the discrete-time state space model
\begin{subequations}
\begin{align}
    \xx_{k+1} &= A^i \xx_k + \ff^i + \ww_k\;,\\
    \yy_{k} &= C \xx_k + \vv_k\;,\\
    \ww_k &\in \mathcal{W}\;,\ \vv_k \in \mathcal{V}\;.
\end{align}
\end{subequations}

\begin{figure}
    \centering
        \input{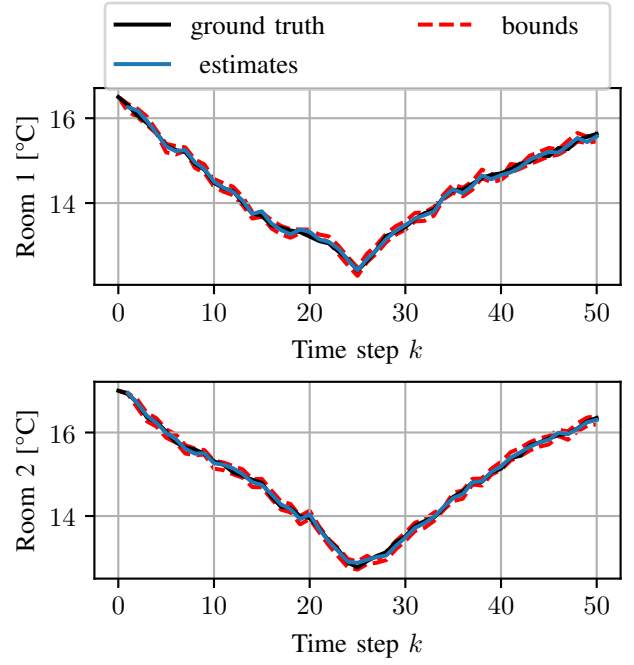}
        \caption{Comparing the ground truth with the optimal MHE estimate and the interval bounds of the feasible sets for the two-room thermal system in Section~\ref{sec:thermal_example}.}
        \label{fig:single_estimator_example}
\end{figure}
\begin{figure}                  
    \input{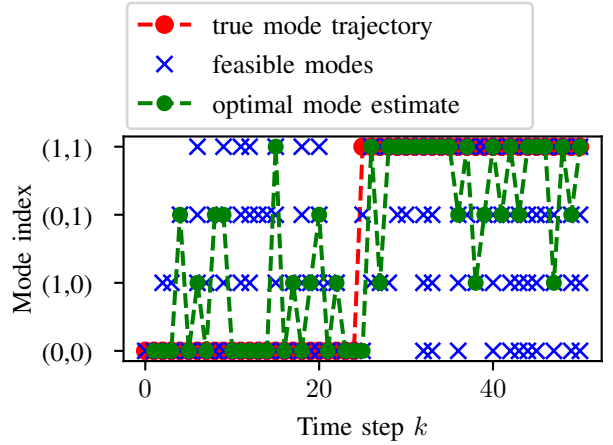}
    \caption{Comparing the feasible modes of the MHE estimates with the true mode and the optimal mode estimates.}
    \label{fig:single_estimator_example_modes}
\end{figure}

Figure~\ref{fig:single_estimator_example} plots the optimal MHE estimate and interval bounds of the currently feasible set in a full moving-horizon simulation using measurement model $C=\eye_2$, horizon length $N=30$, disturbance bounds $\mathcal{W}=\clint{-0.15}{0.15}^2$, and noise bounds $\mathcal{V}=\clint{-0.1}{0.1}^2$. Additionally, Figure~\ref{fig:single_estimator_example_modes} plots the feasible modes for the current time of each call of the estimator. To periodically reduce the complexity of the feasible sets, a bounding box of the feasible states and modes was taken every 10 steps. 

\begin{figure}
    \centering
    \input{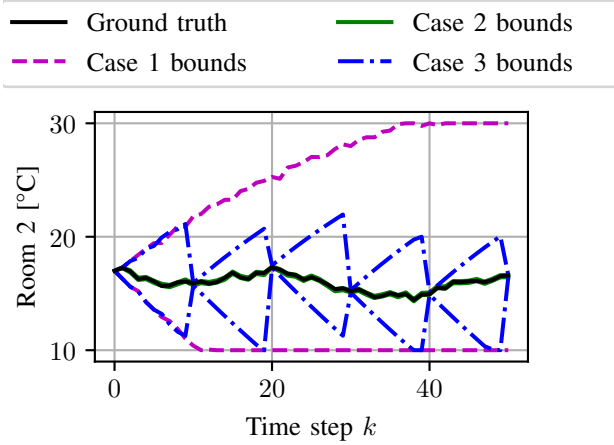}
    \caption{Comparison of three measurement models for the two-room thermal system, plotting both the ground truth temperature of the second room in black and the interval bounds of the feasible estimate sets in colors. Case 1 (magenta) only measures the first room, Case 2 measures both rooms, and Case 3 periodically allows measurement of the second room.}
    \label{fig:compare_meas_models}
\end{figure}

Now, consider three different measurement models for this system. Case 1 uses $C_1 = \begin{bmatrix} 1 & 0 \end{bmatrix}$ where only the first room is measured, whereas Case 2 uses $C_2 = \eye_2$ where both rooms are measured. Lastly considered will be Case 3, where $C_1$ is used for most time steps but using $C_2$ every 10 time steps for periodic measurement of both rooms. Figure~\ref{fig:compare_meas_models} compares these measurement models. The system is not fully observable when the second room is not being measured, and so in Case 1 the interval bounds of the feasible states grow until they reach the entire state space, around time step $k=40$. For Case 2 where both rooms are measured, the interval bounds stay very tight around the actual true states. For Case 3 when measurements of the second room are only periodically available, the interval bounds of the feasible estimate grow significantly in-between samples of the second room, and shrink down when the second room is measured. 

As a point of comparison against the existing approach in the literature, this PWA system can also be written in the MLD form with states $\xx_k$, auxiliary continuous states $\qq_k^i$ for $i \in \{1, \dots, n_m\}$, and auxiliary binary states $\llambda_k \in \{0,1\}^{n_m}$ again corresponding to the modes. The MLD dynamics and constraints are written as
\begin{subequations}
\begin{align}
    \xx_{k+1} &= \sum_{i = 1}^{n_m} \qq_k^i\;,\\
    \forall i &\in \{1, \dots, n_m\} : \\
    &\quad \qq_k^i \leq \mathbf{U}^i \lambda_{k,i}\;,\ \qq_k^i \geq \mathbf{L}^i \lambda_{k,i}\;,\\
    &\quad \qq_k^i \leq A^i \xx_k + \ff^i - \mathbf{L}^i (1-\lambda_{k,i})\;,\\
    &\quad \qq_k^i \geq A^i \xx_k + \ff^i - \mathbf{U}^i (1-\lambda_{k,i})\;,
\end{align}
\end{subequations}
where $\mathbf{U}^i = \max_{\xx\in\mathcal{S}} A^i \xx + \ff^i$, $\mathbf{L}^i = \min_{\xx\in\mathcal{S}} A^i \xx + \ff^i$, are used in a big-M disjunction over the dynamic modes to enforce $\qq_k^i = \lambda_i \xx_{k+1}$. The matrices are concatenated to write the model as
\begin{subequations}
\begin{gather}
    \xx_{k+1} = A_x \xx_k + B_q \qq_k + B_\lambda \llambda_k + \mathbf{b}_\textit{aff} \;,\\
    E_x \xx_k + E_q \qq_k + B_\lambda \llambda_k \leq \mathbf{e}_\textit{aff}\;,
\end{gather}
\end{subequations}
whose graph of function (lifted with mode indicators) can be constructed as
\begin{subequations}
\begin{align}
    \tilde{\Psi}_\textit{MLD}' &\leftarrow \begin{bmatrix}
        \eye \\ \zeros \\ A_x \\ E_x 
    \end{bmatrix} \mathcal{S} \oplus \begin{bmatrix}
        \zeros \\ \zeros \\ B_q \\ E_q
    \end{bmatrix} \mathcal{Q} \oplus \begin{bmatrix}
        \zeros \\ \eye \\ B_\lambda \\ E_\lambda
    \end{bmatrix} \bar{\mathcal{I}}+\begin{bmatrix}
        \zeros \\ \zeros \\ \mathbf{b}_\textit{aff} \\ \zeros 
    \end{bmatrix}\;,\\
    \tilde{\Psi}_\textit{MLD}' &\leftarrow \tilde{\Psi}_\textit{MLD}' \cap_{\begin{bmatrix} \zeros & \zeros & \zeros & \eye \end{bmatrix}} \mathcal{H} \;,\\
    \tilde{\Psi}_\textit{MLD}' &\leftarrow \begin{bmatrix}
        \eye &\zeros &\zeros &\zeros \\
        \zeros &\eye &\zeros &\zeros \\
        \zeros &\zeros &\eye &\zeros
    \end{bmatrix} \tilde{\Psi}_\textit{MLD}'\;,
\end{align}
\end{subequations}
where $\mathcal{Q}$ is a zonotope representing the interval bounds $\clint{\mathbf{L}}{\mathbf{U}}$, $\bar{\mathcal{I}}=\{\llambda \in \{0,1\}^{n_m} \mid \sum_i \lambda_i = 1\}$, and $\mathcal{H} = \{\mathbf{h} \mid \mathbf{h} \leq \mathbf{e}_\textit{aff}\}$. 

For this problem, the PWA approach produces a graph of function with hybrid zonotope complexity $(16, 4, 9)$, whereas the MLD description produces one with complexity $(42, 4, 33)$, due to the presence of the auxiliary variables and the halfspace constraints. Once $\tilde{\Psi}_\textit{MLD}'$ is constructed, then it can be used in the MHE formulation in Section~\ref{sec:methods} to formulate an MHE problem, using hybrid zonotopes, equivalent to the problem formulation in~\cite{FerrariTrecate2002_hybridMHE}.

To analyze the run-times of this MHE using both the PWA and MLD approaches, timed trials were performed over various horizon lengths $N$, with 100 trials in each case. For these timed trials, the initial condition set $\mathcal{X}_0$ was selected as an interval represented by a zonotope (which is the lowest possible complexity, other than knowing the initial condition exactly), and the initial mode was totally unknown (i.e., $\tilde{\mathcal{X}}_0 = \mathcal{X}_0 \times \mathcal{I}$). No warm-starting or arrival costs were utilized. Statistical data for the time taken to solve the MHE for each horizon length is provided in Table~\ref{tab:HVAC_time_trials}. Disparity in the min/max times is due to non-determinism in the solver. In all cases, the PWA approach solved significantly faster than the MLD approach. This is partly because of the reduced complexity, but also because the convex relaxation of the PWA representation is tighter than the MLD representation, as was shown generally (not specific to this example) in~\cite{robbins2026hybrid}.

\begin{table}
    \caption{Times to solve the MHE problem over 100 trials for the two-room thermal system over varying horizon lengths, solved using both PWA and MLD approaches.}
    \label{tab:HVAC_time_trials}
    \centering
    \begin{tabular}{c|c|c|c}
        \toprule
        $N$ & \makecell{min (ms) \\ PWA / MLD} & \makecell{max (ms) \\ PWA / MLD} & \makecell{med (ms) \\ PWA / MLD} \\
        \hline
        5 & 2.5 / 24.9 & 4.3 / 34.9 & 2.8 / 30.4 \\ 
        \hline
        10 & 4.4 / 191.2 & 6.6 / 375.0 & 5.1 / 210.5 \\ 
        \hline
        20 & 9.2 / 378.8 & 18.5 / 1407 & 10.3 / 482.7 \\ 
        \hline
        30 & 14.1 / 636.4 & 30.9 / 2506 & 15.8 / 747.8 \\ 
        \hline
        40 & 20.0 / 873.8 & 57.3 / 5787 & 22.0 / 2022 \\ 
        \hline
        50 & 26.1 / 1026 & 63.5 / 8864 & 29.2 / 2629 \\
        \bottomrule
    \end{tabular}
\end{table}

\begin{figure}
    \centering
    \input{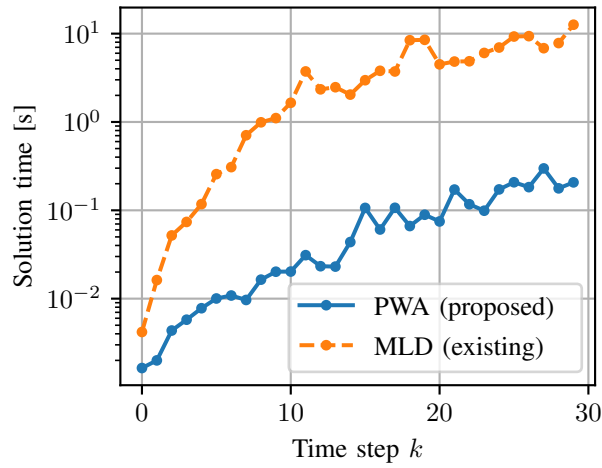}
    \caption{Comparing the solution times of the PWA and MLD approaches in a single moving horizon simulation with $N=10$, propagating estimated sets between calls of MHE without complexity reduction.}
    \label{fig:comp_times_PWA_MLD}
\end{figure}

Because the initial condition was taken as an interval and not a propagated estimate from a previous MHE, these times reported in Table~\ref{tab:HVAC_time_trials} are lower than the actual times needed to solve the problem in practice. Figure~\ref{fig:comp_times_PWA_MLD} plots the actual solution times for a single trial over a simulation of 30 steps with a horizon of $N=10$ and without complexity reduction, to see how the solution time increases as the complexity increases. Both approaches see the complexity increase over time cause an increase in the solution time, but it is far more dramatic for the MLD approach whose graph of function has higher complexity. 
The PWA approach saw a maximum solve time of 298.0ms at time $k=27$, whereas the MLD approach saw a maximum solve time of 12619ms at $k=29$. After 30 time steps, the complexity of the PWA problem was $(6630, 1020, 4845)$, while the MLD problem had complexity $(13260, 1020, 10965)$. 

\section{Conclusion}
\label{sec:conclusion}
This paper proposes a method to formulate and solve moving horizon estimation problems for hybrid systems. Hybrid zonotopes are used to represent the feasible space and encode the set of states consistent with the dynamics and measurements. 
The approach is demonstrated on two numerical examples, a vehicle changing driving lanes and a two-room thermal system with uncertain discrete modes. For the two-room thermal system, solution times are compared for the proposed approach and the MLD approach standard in the literature, and the proposed approach was found to solve much faster due to its lower complexity and tighter convex relaxations. Future work will explore additional methods for periodically reducing the set representation complexity between estimation iterations.

\bibliography{bibitems}
\bibliographystyle{ieeetr}

\end{document}